\documentclass[12pt]{article}

\usepackage{graphicx,psfrag,epsf}
\usepackage{enumerate}
\usepackage{natbib}
\usepackage{mathtools}
\usepackage{booktabs}
\usepackage{color}
\usepackage[english]{babel} 
\usepackage[protrusion=true,expansion=true]{microtype} 
\usepackage{amsmath,amsfonts,amsthm}
\usepackage{amssymb}
\usepackage{chngcntr}
\usepackage[toc,page]{appendix}

\usepackage[colorlinks, linkcolor=blue, anchorcolor=blue, citecolor=blue]{hyperref}

\usepackage{array}
\usepackage{booktabs} 
\usepackage{natbib}
\usepackage{setspace}

\usepackage{soul}
\usepackage{textcomp}
\usepackage{multirow}
\usepackage{enumitem}
\usepackage{microtype} 
\usepackage[font = small,labelfont=bf,textfont=it]{subcaption}
\usepackage{color}
\usepackage{tabularx}
\usepackage[font = small,labelfont=bf,textfont=it]{caption} 
\usepackage{footnote}
\usepackage{algorithm}
\usepackage[noend]{algpseudocode}
\usepackage{etoolbox}

\algblock{ParFor}{EndParFor}
\algnewcommand\algorithmicparfor{\textbf{for}}
\algnewcommand\algorithmicpardo{\textbf{do\ parallel}}
\algnewcommand\algorithmicendparfor{\textbf{end\ parallel\ for}}
\algrenewtext{ParFor}[1]{\algorithmicparfor\ #1\ \algorithmicpardo}
\algrenewtext{EndParFor}{\algorithmicendparfor}

\makeatletter
\def\BState{\State\hskip-\ALG@thistlm}

\newcommand{\distas}[1]{\mathbin{\overset{#1}{\kern\z@\sim}}}%
\newcommand{\bm}[1]{\mathbf{#1}}
\newsavebox{\mybox}\newsavebox{\mysim}
\newcommand{\distras}[1]{%
  \savebox{\mybox}{\hbox{\kern3pt$\scriptstyle#1$\kern3pt}}%
  \savebox{\mysim}{\hbox{$\sim$}}%
  \mathbin{\overset{#1}{\kern\z@\resizebox{\wd\mybox}{\ht\mysim}{$\sim$}}}%
}

\newcommand{\be}{\begin{equation}}
\newcommand{\ee}{\end{equation}}
\newcommand{\bi}{\begin{itemize}}
\newcommand{\ei}{\end{itemize}}
\newcommand{\ben}{\begin{enumerate}}
\newcommand{\een}{\end{enumerate}}
\newcommand{\stb}{\State $\bullet$ \;}

\newcolumntype{K}[1]{>{\centering\arraybackslash}p{#1}}
\allowdisplaybreaks
\DeclareMathOperator*{\argmin}{\arg\!\min}
\makeatother

\let\oldbibliography\thebibliography
\renewcommand{\thebibliography}[1]{\oldbibliography{#1}
\setlength{\itemsep}{0pt}} 

\newcommand{\blind}{1}

\patchcmd{\footnotemark}{\stepcounter{footnote}}{\refstepcounter{footnote}}{}{}

\begin{document}

\def\spacingset#1{\renewcommand{\baselinestretch}%
{#1}\small\normalsize} \spacingset{1}

\if1\blind
{
  \title{\bf A calibration-free method for biosensing \\ in cell manufacturing}
  \small
  \author{Jialei Chen\thanks{H. Milton Stewart School of Industrial and Systems Engineering, Georgia Institute of Technology} \thanks{Georgia Tech Manufacturing Institute, Georgia Institute of Technology}~,  Zhaonan Liu$^{\dagger}$\thanks{School of Materials Science and Engineering, Georgia Institute of Technology}~, Kan Wang$^{\dagger}$, Chen Jiang$^{\dagger}$$^{\ddag}$, \\
  Chuck Zhang$^{*}$$^{\dagger}$, 
  and Ben Wang$^{*}$$^{\dagger}$$^{\ddag}$ \thanks{This work is supported by National Science Foundation CMaT ERC (NSF EEC-1648035).
  }\hspace{.2cm}\\
}
  \maketitle
} \fi

\if0\blind
{
  \bigskip
  \bigskip
  \bigskip
  \begin{center}
    {\LARGE\bf  A calibration-free method for biosensing \\ \vspace{.3cm} in cell manufacturing}
\end{center}
  \medskip
} \fi

\bigskip

\vspace{-0.5cm}
\begin{abstract}
Chimeric antigen receptor T cell therapy has demonstrated innovative therapeutic effectiveness in fighting cancers; however, it is extremely expensive due to the intrinsic patient-to-patient variability in cell manufacturing.
We propose in this work a novel calibration-free statistical framework to effectively recover critical quality attributes under the patient-to-patient variability.
Specifically, we model this variability via a patient-specific calibration parameter, and use readings from multiple biosensors to construct a patient-invariance statistic, thereby alleviating the effect of the calibration parameter. A carefully formulated optimization problem and an algorithmic framework are presented to find the best patient-invariance statistic and the model parameters. Using the patient-invariance statistic, we can recover the critical quality attribute of interest, free from the calibration parameter. We demonstrate improvements of the proposed calibration-free method in different simulation experiments. In the cell manufacturing case study, our method not only effectively recovers viable cell concentration for monitoring, but also reveals insights for the cell manufacturing process.

\end{abstract}

\noindent%
\vfill

\newpage
\spacingset{1.45} 

\section{Introduction} \label{sec:intro}

\begin{figure}
\centering
\includegraphics[width=0.70\textwidth]{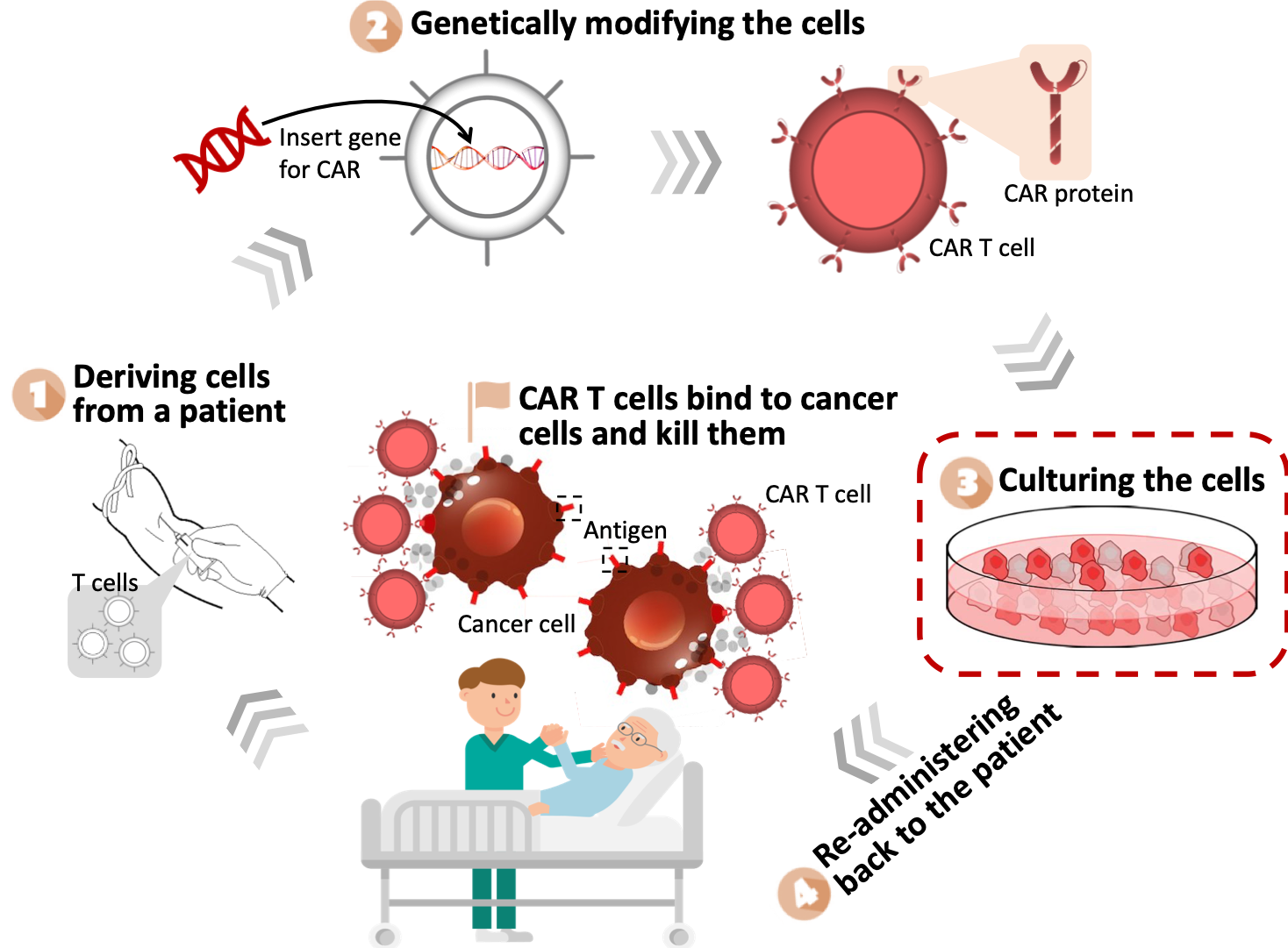}
\caption{\label{fig:CART} An illustration of the four steps in a typical CAR T cell therapy. This work focuses on the cell culturing (or cell manufacturing) step, i.e., step 3.}
\end{figure}

Cell therapy is one of the most promising new treatment approaches over the last decades, demonstrating great potential in treating cancers, including leukemia and lymphoma \citep{kim2009stem,yin2017keys}. Among those therapies, chimeric antigen receptor (CAR) T cell therapy \citep{bonifant2016toxicity,june2018car}, involving the reprogramming of a patient’s T cells to effectively target and attack tumor cells, has shown innovative therapeutic effects in clinical trials, leading to a recent approval (i.e., the treatment of CD19+ hematological malignancies, see \citealp{prasad2018tisagenlecleucel}) by FDA as a new cancer treatment modality. As illustrated in Figure \ref{fig:CART}, a typical CAR T cell therapy involves four steps -- deriving cells from a patient, genetically modifying the cells, culturing the cells, and re-administering back to the patient. 
With increasingly mature gene modification technology, more and more researchers focus on the culturing step (i.e., the red box in Figure \ref{fig:CART}), where the goal is to substantially increase the cell amount from a small batch to one dose for delivery to the patient. However, a key challenge is the intrinsic patient-to-patient variability in the starting material, i.e., cells derived from different patients vary in their viabilities, acceptance rates of genetic modification, and reactions to culture media \citep{hinrichs2013reassessing}. These variabilities introduce difficulties in cell culturing scale-up (i.e., cell manufacturing), and therefore, the current CAR T cell therapy is hindered by low scalability, labor-intensive processes, and extremely high cost \citep{harrison2019chimeric}. To achieve high quality and acceptable vein-to-vein cost, we present in this work a statistical framework for online monitoring in cell manufacturing, which can alleviate the negative effect of the intrinsic patient-to-patient variability.

There are two reasons why a new statistical method is needed for monitoring critical quality attributes (exampled by the cell concentration) in cell manufacturing. Firstly, a \textit{direct} measurement method for cell concentrations is not suitable in cell manufacturing  \citep{slouka2016novel}. Such a method typically requires experienced technicians to collect culture media, take microscopic images, and perform computation via an image-based software (e.g., ImageJ, see \citealp{collins2007imagej}). Therefore, it is labor-intensive, time-consuming, and may introduce contamination to the culture media. Furthermore, the direct measurement method is oftentimes \textit{destructive} -- the collected cells would be killed for taking microscopic images. Secondly, while there are \textit{non-destructive} sensors available, these sensors need to be calibrated due to the unknown parameters in the sensing relationship \citep{pan20193d}. For example, impedance sensors  (adopted in this work, see Figure \ref{fig:biosensor}), which measure the dielectric relaxation of cell suspension, can be used to effectively estimate cell concentrations \textit{after} the calibration of unknown electrical attributes, e.g., permittivity \citep{gheorghiu1998monitoring} and resistivity \citep{goh2010impedance}. However, 
due to the patient-to-patient variability, those electrical attributes not only are \textit{unknown} but also \textit{vary} among different patients, leading to difficulties in recovering cell concentrations from sensor readings.

We introduce in this work a calibration-free statistical method for online monitoring in cell manufacturing. Specifically, the intrinsic patient-to-patient variability is modeled by a patient-specific calibration parameter. 
We propose to use multiple sensor readings to construct a patient-\textit{invariance} statistic, where a transformation is adopted to isolate and alleviate the effect of the calibration parameter. The constructed invariance statistic is then used to model the critical quality attribute of interest. In the \textit{training} stage, we use the historical data to estimate a transformation and  model parameters via a carefully formulated optimization problem, rather than estimate the calibration parameter as in the standard calibration problem \citep{KO2001,tuo2015efficient}. In the \textit{monitoring} stage, we use the online sensor readings to recover the underlying critical quality attribute through the patient-invariance statistic, \textit{free} from the calibration parameter. We demonstrate improvements of the proposed calibration-free method in both simulation experiments and a real-world case study of monitoring viable cell concentrations in cell manufacturing. The proposed approach provides an effective way to monitor cell manufacturing, and therefore, reduces the cost for the promising CAR T cell therapy in treating cancers.

The remaining part of the article is organized as follows. In Section \ref{sec:probdefin}, we formulate the biosensing problem in cell manufacturing, with an emphasis on its challenging aspects. 
In Section \ref{sec:method}, we present the proposed calibration-free method. A detailed simulation study and a real-world cell manufacturing case study are conducted in Sections \ref{sec:simstudy} and \ref{sec:casestudy}, respectively. We conclude this work with future directions in Section \ref{sec:conclusion}.

\section{Biosensing in cell manufacturing} 
\label{sec:probdefin}

We first describe the biosensing problem of recovering the Viable Cell Concentration (VCC) in cell manufacturing. We then discuss the key challenge -- the patient-to-patient variability, and related works.

\subsection{Impedance-based biosensing}
\label{sec:impbiosens}

\begin{figure}
\centering
\includegraphics[width=0.75\textwidth]{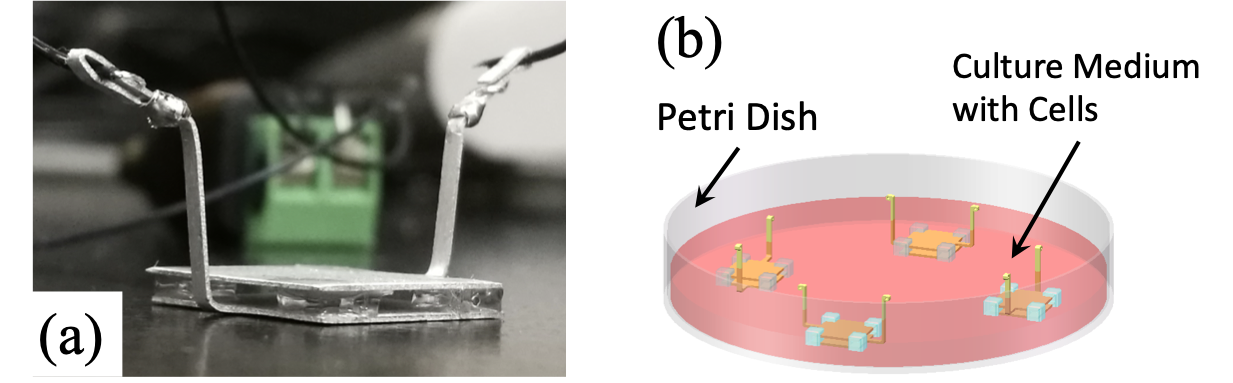}
\caption{\label{fig:biosensor} An illustration of the adopted impedance-based biosensors in the cell manufacturing application: (a) shows a photo of the biosensor design and (b) shows the biosensing setup.}
\end{figure}

As discussed in Section \ref{sec:intro}, the goal is to monitor VCC in cell manufacturing, thereby reducing the cost of the CAR T cell therapy.
One state-of-the-art approach is to use biosensors to measure impedance signals, as indicators for the VCC of interest \citep{gheorghiu1998monitoring, pan20193d}.
As illustrated in Figure \ref{fig:biosensor}, we adopt impedance-based biosensors with a facing-electrode (FE) design \citep{miura2019computer}:
Our FE biosensor consists of a pair of parallel-plate electrodes and silicone at four corners to maintain a gap between them; it would be soaked in media to monitor floating cells in between the electrodes.

With the adopted biosensors, we need a biosensing method to recover VCCs from impedance readings. 
From physical knowledge, we know that the impedance reading between the two electrodes reflects the cell amount due to the capacitive property of viable cell membranes \citep{schwan1957electrical}. The sensing relationship is denoted by
\begin{align}
    y = f(x, \theta, \eta),
    \label{eq:sensing}
\end{align}
where $y$ is the impedance reading, $x$ is the underlying VCC of interest, $\theta$ denotes the sensor geometry, e.g., the gap width, and $\eta$ models the underlying electrical attributes, e.g., permittivity and resistivity. 
Here, the relationship $f(\cdot)$ is \textit{unknown} due to the dynamic interaction between cells and biosensors, and extremely \textit{challenging} to simulate via computer codes considering the micro size of cells.
In the proposed calibration-free method, we first learn the sensing relationship $f(\cdot)$ from the historical data of different patients (i.e., the ``training" stage), and then conduct online inference of the VCC using the impedance readings for new patients (i.e., the ``monitoring" stage).

\subsection{Patient-to-patient variability} \label{sec:p2pvar}

The key challenge in biosensing is that the electrical attributes $\eta$ are  \textit{unknown} in both training and monitoring and \textit{different} for each patient (also see an illustrating gravity application in Section \ref{sec:gapp}). 
Note that it is impractical to compute $\eta$ for a patient from first principle since it represents the intrinsic properties from genetic material of the patient. One popular way is to model $\eta$ as a \textit{calibration} parameter, and estimate it from the training dataset $\{y^j, x^j, \theta\}_{j=1}^J$. Existing approaches include Bayesian implementation \citep{KO2001}, maximal  likelihood estimation \citep{joseph2009statistical}, and an interpretative $l_2$ optimization \citep{tuo2015efficient}. 
However, such methods are proposed specifically for data fusion of computer experiments and physical experiments, where  $\eta$ is \textit{available} in the former and a \textit{constant} in the latter. Whereas in biosensing, there is \textit{no} effective computer simulator, and electrical attributes $\eta$ \textit{vary} among physical experimental runs for different patients.
In the literature, this challenge is also related to the functional calibration problem \citep{plumlee2016calibrating,brown2018nonparametric,ezzat2018sequential}, where the calibration parameter $\eta=\eta(x)$ is modeled as a function of the input variables. In the biosensing application, however, the calibration parameter $\eta$ varies among different patients yet is a constant over different input VCCs for each patient.

\begin{table}[]
\centering
\begin{tabular}{l|ccc}
 \toprule
Methods & $f(\cdot)$ & Online $\eta^*$ &  Historical $\eta^j$ \\
\hline\hline
Inverse problem \citep{aster2018parameter}     & Known       & Unknown       &  N.A.            \\
Supervised learning \citep{bishop2006pattern}  & Unknown       & Same     &    Same \\
Calibration \citep{KO2001}   & Unknown       & Unknown      & Known            \\
 \hline 
\textbf{Calibration-free} (proposed)    & Unknown       & Unknown      & Unknown        \\
 \toprule
\end{tabular}
\caption{A comparison of the application scenarios of the proposed calibration-free method and other standard methods in the literature.}
\label{ta:diffMethods}
\end{table}

The biosensing problem is also related to the inverse problem in the literature \citep{aster2018parameter}, where one would estimate both $x$ and $\eta$ via an optimization problem. However, such a method typically assumes that the sensing relationship $f(\cdot)$ is known or can be easily learned with a \textit{complete} data $\{y^j, x^j,\theta,\eta^j\}_{j=1}^n$, whereas, the calibration parameter $\eta$ is  \textit{unknown} even in the training stage in biosensing. 
Furthermore, one may regard $\eta$ as an additional model parameter in the unknown relationship $f(\cdot)$, and adopt a standard supervised learning scheme \citep{bishop2006pattern} for finding both $f(\cdot)$ and $\eta$; this implicitly assumes that $\eta$ is a constant for different patients, which is \textit{not} true in biosensing.
Table \ref{ta:diffMethods} summarizes the related methods discussed;
with extensive efforts in literature search, we have not found a standard method, which can be directly adopted for the biosensing problem in cell manufacturing.

\section{Calibration-free biosensing method} \label{sec:method}

We present the proposed calibration-free method in four parts. First, we discuss the sensing relationship for multiple sensors. We then introduce an invariance statistic to alleviate patient-to-patient variability. In the online monitoring stage, we use the invariance statistic to recover VCCs. In the training stage, we propose a carefully constructed optimization problem and an algorithmic framework to estimate the underlying sensing model.

\subsection{Sensing relationship with multiple sensors}

The key idea of the calibration-free method is to use \textit{multiple} sensors to address the unknown and patient-specific calibration parameter. For the $i$-th sensor, we let $\theta_i$ be its geometry parameter.
Consider first the sensing relationship for a given patient (or experiment). Denote $y_i[t]$ as the scalar impedance reading  (more details in Section \ref{sec:casestudy}) from the $i$-th sensor at experimental time $t$. Following \eqref{eq:sensing}, we model $y_i[t]=f(x[t],\theta_i,\eta)$ via the sensing relationship $f(\cdot):\mathbb{R}^3 \mapsto \mathbb{R}$. Here, $x[t]$ is the VCC at experimental time $t$, and $\eta$ is the calibration parameter. It is important to note that measurements from different sensors $\{y_1[t],y_2[t],\cdots\}$ can be modeled by a \textit{same} calibration parameter $\eta$, featuring the underlying values for electrical attributes of the specific patient's cells. This patient-specific property is the key to ``canceling out" the calibration parameter using multiple sensor readings (see Section \ref{sec:invarstat}).

We then introduce an additional superscript $j$ for different patients:
\begin{align}
y^j_i[t] = f(x^j[t],\theta_i, \eta^j), \quad i=1,2,\cdots, I, \quad t=1,\cdots T,\quad \text{and,} \quad j=1,\cdots, J.
\label{eq:model}
\end{align}
Equation \eqref{eq:model} further layouts our biosensing settings with multiple sensors: (i) We assume the \textit{homogeneity} of VCC, i.e., at given time $t$ and a given patient $j$, the VCC $x^j[t]$ is the same for different senors $\theta_i$ at different locations (see Figure \ref{fig:biosensor} (b)). This is because, in suspension cell manufacturing, the culture media is constantly stirred to ensure the homogeneity of nutrition, and thereby VCC \citep{haycock20113d}. 
(ii) We use the same set of sensors with \textit{known} parameters $\{\theta_i\}_{i=1}^I$ for all $J$ patients and at different experimental time $t$. Those parameters are known from the fabrication process or can be easily measured from the sensors.  Note that the proposed method is also effective for different sets of sensors, as long as those sensor parameters are known -- the same sensor assumption is only for fabrication convenience and notation simplicity.
(iii) Besides for different sensors $\theta_i$, the calibration parameter $\eta^j$ is the \textit{same} among different measurement time $t$. This is because $\eta^j$ models the intrinsic property of the $j$-th patient's cells, 
which typically does not change during cell manufacturing. 
After we clearly layout the above settings, we can then construct the patient-invariance statistic and recover the underlying VCC.

\subsection{Invariance statistic} \label{sec:invarstat}
For notation simplicity, we drop the experimental time $[t]$ and write $\theta_i$ in the subscript in this subsection.  Furthermore, we rewrite \eqref{eq:model} by decomposing $f(\cdot)$ into two parts:
\begin{equation}
y^j_i = f_{\theta_i}(x^j, \eta^j) = \mu_i(x^j) + \delta_i(x^j,\eta^j).
\label{eq:model_simp}
\end{equation}
Here, for a given sensor $\theta_i$, $\mu_i(\cdot):\mathbb{R}\mapsto \mathbb{R}$ models the part of effect of VCC $x$ on impedance reading $y$,  \textit{without} hampered by the patient-specific calibration parameter $\eta$; and $\delta_i(\cdot):\mathbb{R}^2 \mapsto \mathbb{R}$ is the remaining effect of \textit{both} $x$ and $\eta$ on $y$. 
Intuitively speaking, $\mu_i(\cdot)$ can be viewed as the \textit{mean} process of $f(\cdot)$ by plugging in some population average of $\{\eta^j\}_{j=1}^J$, ignoring the patient-to-patient variability; it can also be interpreted as the physical understanding of the sensing relationship. 
In practice, the mean relationship $\mu_i(\cdot)$ is oftentimes known, at least to a certain degree, prior to experimentation according to the domain-specific knowledge (e.g., the known set of basis functions, see Section \ref{sec:casestudy}). On the other hand, $\delta_i(\cdot)$ is the \textit{variability} term, i.e., how patient-to-patient variability affects the impedance reading. Such a term leads to different readings, even when the VCC $x$ is the same. Note that in one of the considered baseline methods  (see Sections \ref{sec:simstudy} and \ref{sec:casestudy}), we ignore $\delta_i(\cdot)$, i.e., assuming the calibration parameter is a constant; this will lead to noticeable errors when estimating $x$. This variability term $\delta_i(\cdot)$ is typically unknown. Such a decomposition of a mean trend and a variability term is widely assumed in different modeling methods (see, e.g., \citealp{guillas2018functional,chen2019function}).

Assume for now $\delta_i(x,\eta)=\delta_i(\eta)$, which suggests that the mean relationship $\mu_i(\cdot)$ extracts  all the dependency of $x$ on $y$ (further discussion in Section \ref{sec:paraestim}). In other words, $f_i(x,\eta)$ is assumed to be \textit{separable} for each sensor $\theta_i$:
\begin{align}
y^j_i = \mu_i(x^j) + \delta_i(\eta^j).
\label{eq:model_sep}
\end{align}

Now, we construct a statistic $F$ which is invariant to the calibration parameter $\eta$. To gain intuition, consider the following illustrating example with \textit{known}  $\delta_i(\eta)=\theta_i\eta$ for $i=1,2$ (see the illustration application in Section \ref{sec:gapp}). If we take a log-transformation to the variability term $\log \delta_i(\eta^j) = \log\theta_i+\log\eta^j$, the effect of the calibration parameter $\eta^j$ is further separated from $\theta_i$. Therefore, by subtracting the (log-transformed) variability at different sensors, one can obtain an invariance statistic $F=\log \delta_1(\eta^j)- \log \delta_2(\eta^j) = \log\theta_1+\log\eta^j-(\log\theta_2+\log\eta^j) =\log(\theta_1)-\log(\theta_2)$. Note that we incorporate the patient-\textit{specific} property of the calibration parameter when constructing the invariance statistic.

Following the above intuition yet with the \textit{unknown} variability term $\delta_i(\cdot)$, we construct the following statistic, via a linear combination of the transformed $\delta_i(\cdot)$:
\begin{align}
F(\eta^j) = \sum_{i=1}^I c_i \mathcal{F} [\delta_i(\eta^j)] =  \sum_{i=1}^I c_i \mathcal{F} [y_i
^j - \mu_i(x^j)],
\label{eq:aff_comb}
\end{align}
for patient $j$ with $\eta^j$. Here, $c_1,\cdots, c_I$ are pre-defined combination coefficients, and $\sum_i c_i=0$.
With a properly selected transformation $\mathcal{F}[\cdot]:\mathbb{R}\mapsto \mathbb{R}$, \eqref{eq:aff_comb} gives the target \textit{invariance} statistic $F=F(\eta^j)$.
Note that here we adapt a general transformation $\mathcal{F}[\cdot]$, instead of the specific log-transformation in the above example. The transformation $\mathcal{F}[\cdot]$ would be selected so that the dependency of the invariance statistic $F$ to $\eta^j$ is minimal; a detail estimation method for $\mathcal{F}[\cdot]$ will be discussed in Section \ref{sec:paraestim}.

It is important to note that we reconstruct the sensing model from \eqref{eq:model} to \eqref{eq:aff_comb} via the proposed invariance statistic. This is again due to the key challenge of patient-to-patient variability. Consider first using \eqref{eq:model} for VCC recovery (also see the discussion in Section \ref{sec:p2pvar}). Due to the unknown and patient-specific calibration parameter $\eta^j$, it is challenging to either learn a sensing model from training data or recover VCCs for a new patient. However, the new model \eqref{eq:aff_comb} contains only the invariance statistic, and is \textit{free} from the calibration parameter. Thanks to the properly selected transformation and the combination (see Section \ref{sec:paraestim}), the invariance statistic would be approximately a constant for different patients. Therefore, our new model \eqref{eq:aff_comb} allows an effective estimation of the sensing relationship (only the mean part needed) similar to the standard calibration problem with a \textit{constant} calibration parameter \citep{tuo2015efficient}, and then a calibration-free recovery of the VCC of interest.

 \subsection{Online calibration-free recovery} \label{sec:recovery}

We present next the method for recovering the VCC of interest $x^*$, in the \textit{online monitoring} stage for a new patient denoted by $*$.
At any time $t$, the sensor reading is denoted as $\mathcal{D}_{\rm monitor} = \{y_i^*,\theta_i\}_{i=1}^I$ along with the unknown calibration parameter $\eta^*$. Assume for now the mean sensing relationship  $\mu_i(\cdot)$ and the transformation $\mathcal{F}[\cdot]$ are known (see Section \ref{sec:paraestim} for the estimation). We adopt the new sensing model \eqref{eq:aff_comb} with the invariance statistic
\begin{align}
   F(\eta^*)= \sum_{i=1}^I c_i \mathcal{F} \left[y_i^*-\mu_i(x^*)\right],
\end{align}
where $x^*$ is the target VCC. Note that the computed $F(\eta^*)$ in online monitoring is also \textit{invariant} to the calibration parameter $\eta^*$.  
Therefore, the VCC of interest $x^*$ can be recovered by minimizing the squared difference between the computed value and the underlying value $\bar{F}$ (see Section \ref{sec:paraestim} for the estimation):
\begin{align}
    \hat{x}^* = \argmin_{x^*} \left(\sum_{i=1}^I c_i \mathcal{F} \left[y_i^*-\mu_i(x^*)\right] - \bar{F} \right)^2. 
    \label{eq:estimatex}
\end{align}

In the cell manufacturing application, we are interested in recovering a VCC curve $\hat{x}^*[t]$ over the whole manufacturing period $t=1,\cdots, T$. To this end, we perform optimization \eqref{eq:estimatex} for $T$ times corresponding to each experimental time $t$. Note that here we have not incorporated the time-dependency (or smoothness) of the recovered function $\hat{x}^*[t]$ in online recovering; one can use postprocessing methods or directly model $x^*[t]$ via a parametric form in the optimization \eqref{eq:estimatex}. Readers are referred to functional data analysis literature \citep{ramsay2004functional,wang2016functional} for more discussion.

\subsection{Parameter estimation} \label{sec:paraestim}
We estimate the unknown transformation $\mathcal{F}[\cdot]$, and the physical relationship $\mu_i(\cdot)$ using the training data $\mathcal{D}_{\rm train} = {\{y_i^j[t],x^j[t],\theta_i\}_{t=1}^T}_{j=1}^J$ at hand.
In our implementation, the transformation $\mathcal{F}[\cdot]$ is paramterized by the Box-Cox transformation \citep{box1964analysis,sakia1992box} 
\begin{equation}
\mathcal{F}_\lambda[z] = 
\left\{ \begin{matrix}
\log(z) &\quad \text{if}\quad \lambda=0 \\
\frac{\left(z^\lambda-1\right)}{\lambda} &\quad \text{if}\quad \lambda\neq 0
\end{matrix}\right..
\label{eq:BoxCox}
\end{equation}
Note that the log-transformation in the above example is a special case in the Box-Cox transformation. Here, the Box-Cox transformation contains an unknown parameter $\lambda$. A two-parameter Box–Cox or Yeo–Johnson transformation \citep{yeo2000new} can also be used if the data are not restricted to be nonnegative. Furthermore, in this article, we will focus on the parametric transformation \eqref{eq:BoxCox}, but our proposed method are general and can be applied to non-parametric cases.

As for the physical relationship $\mu_i(\cdot)$, we adopt the following basis decomposition:
\begin{align}
     \mu_{i,\boldsymbol\beta}(x) = \mu_{\boldsymbol\beta} (x,\theta_i) = \sum_{k=1}^K \beta_k \phi_k(x,\theta_i).
\end{align}
Here, $\phi_k(\cdot);\; k=1,2,\cdots, K$ are the pre-defined basis functions and $\boldsymbol\beta  = [\beta_1,\cdots, \beta_K]^T$ denotes the vector of corresponding coefficients. Such a set of basis $\{\phi_k(\cdot)\}_{k=1}^K$ is selected by the physical knowledge of the cell manufacturing system or the observation from data.

Meanwhile, to account for the separable assumption $\delta_i(x,\eta) \approx \delta_i(\eta)$ in Section \ref{sec:invarstat}, we introduce slack variables $\boldsymbol\Delta = {\{\Delta_i^j\}_{i=1}^I}_{j=1}^J$ to account for the ``goodness" of the assumption (more discussion below). Furthermore, the underlying value for the best invariance statistic $\bar{F}$ is also unknown and need to be estimated from data (see Section \ref{sec:recovery}).

We propose to estimate the unknown parameters $\{\lambda, \boldsymbol\beta, \bar{F}, \boldsymbol\Delta \}$ via the following optimization problem with two penalization terms:
\begin{align}
\begin{split}
  \min_{\lambda,\boldsymbol\beta,\bar{F}, \boldsymbol\Delta} l_\alpha(\lambda,\boldsymbol\beta,\bar{F}, \boldsymbol\Delta ) =   \min_{\lambda,\boldsymbol\beta,\bar{F}, \boldsymbol\Delta} & \sum_{t,j}\left[\sum_{i=1}^I c_i\mathcal{F}_\lambda[y_i^j[t] - \mu_{\boldsymbol\beta}(x^j[t],\theta_i)]-\bar{F}\right]^2 \\
   & + \alpha_1\sum_{i,j,t}  \left[|y^j_i[t]-\mu_{\boldsymbol\beta}(x^j[t],\theta_i)|_1-\Delta_i^j\right]^2 + \alpha_2 |\boldsymbol\beta|_1. 
    \end{split}
\label{eq:paraestimate}
\end{align}
Here, $\alpha_1$ and $\alpha_2$ are two penalization coefficients, and $|\cdot|_1$ denotes the vector $l_1$ norm.

The main objective term (i.e., the first term) in \eqref{eq:paraestimate} is for achieving the best patient-invariance property of the constructed statistic $F$. Specifically, we minimize the mean-squared error (MSE) of its underlying truth $\bar{F}$ and the computed value from data. This is equivalent to modeling the patient-invariance statistic $F^j$ for each patient as i.i.d.~random draws from a normal distribution $\mathcal{N}(\bar{F}, \sigma^2)$ with mean $\bar{F}$ and variance $\sigma^2$.
Moreover, the first penalization term in \eqref{eq:paraestimate} is for the separable assumption $\delta_i(x,\eta) \approx \delta_i(\eta)$. Here, we minimize the corresponding MSE of the set $\{\delta_i(x^j[t],\eta^j)\}_{t=1}^T$ to the underlying truth $\Delta_i^j$, for each sensor $i$ and patient $j$.
Similarly, this can also be viewed as modeling $\delta_i(\cdot, \eta^j)$ via i.i.d.~normal random variables; the corresponding penalization $\alpha_1$ can then be interpreted as the ratio between the variances of the two normal distributions.
Finally, the second penalization term  $\alpha_2 |\beta|_1$ is for basis selection, similar to the widely used LASSO method in the literature \citep{tibshirani1996regression,donoho2006most}. 
This is because, in the cell manufacturing application, one would only have an intuitive understanding of the sensing relationship; we will collect a set of basis functions from experience and select the suitable ones via this penalization.

From the duality of the optimization problem, \eqref{eq:paraestimate} can also be viewed as unpenalized log-likelihood combining both normal random variables with a sparsity constraint $\|\boldsymbol\beta\|_1 \leq \gamma$ \citep{boyd2004convex}. The parameter $\alpha_1$ sets the variance ratio between the two random variables, and $\alpha_2$ controls the desired sparsity level as parameter $\gamma$. Since the objective is to obtain a high recovery accuracy of the VCC of interest, $\alpha_1$ and $\alpha_2$ would be tuned using cross-validation techniques \citep{friedman2001elements}. 

\begin{algorithm}[t]
\caption{The BCD algorithm for parameter estimation \eqref{eq:paraestimate} }\label{BCD}
\label{alg:BCD}
\begin{algorithmic}[1]
\small
\stb Set initial values $\lambda \leftarrow 0$ and $\boldsymbol\beta \leftarrow \bm{1}_{K}$
\stb Set $I\times J\times T\times K$ tensor  $\boldsymbol\Phi$ with each element $\Phi_{ijtk}  \leftarrow  \phi_k(x^j[t],\theta_i)$
\Repeat\\
\quad \quad \quad \underline{Optimizing $F$ and $\boldsymbol\Delta$}: 
\stb Set $I\times J\times T$ tensor $\bm{D}$ with each element $D_{ijt} \leftarrow y_i^j[t] - \sum_k \Phi_{ijtk} \beta_k$
\stb Update $\bar{F} \leftarrow \sum_{ijt} c_i \mathcal{F}_\lambda [D_{ijt} ]/J/T$
\stb Update $\Delta_i^j \leftarrow \sum_{t} |c_i \mathcal{F}_\lambda [D_{ijt} |_1/T$ for $i=1,\cdots I$ and $j=1,\cdots J$\\
\quad \quad \quad \underline{Optimizing $\lambda$}:
\stb Set $l_0(\lambda,\boldsymbol\beta,\bar{F}, \boldsymbol\Delta ) \leftarrow  l_\alpha(\lambda,\boldsymbol\beta,\bar{F}, \boldsymbol\Delta )$ with $\alpha_1 = \alpha_2= 0$
\stb Update $\lambda \leftarrow \argmin_{\lambda} l_\alpha(\lambda,\boldsymbol\beta,\bar{F}, \boldsymbol\Delta )$\\
\quad \quad \quad \underline{Optimizing $\boldsymbol{\beta}$}:
\stb Update $\boldsymbol\beta \leftarrow \argmin_{\boldsymbol\beta} l_\alpha(\lambda,\boldsymbol\beta,\bar{F}, \boldsymbol\Delta )$ with the L-BFGS method
\Until{$\lambda$, $\bar{F}$, $\boldsymbol\Delta$ and $\boldsymbol{\beta}$ converge}
\stb \Return $\lambda$, $\bar{F}$, $\boldsymbol\Delta$ and $\boldsymbol{\beta}$
\normalsize
\end{algorithmic}
\end{algorithm}

Consider now estimating the parameters $\{\lambda, \boldsymbol\beta, \bar{F}, \boldsymbol\Delta \}$  via optimization \eqref{eq:paraestimate} for fixed $\alpha_1>0$ and $\alpha_2>0$. We propose to use the following Blockwise Coordinate Descent (BCD) optimization algorithm, described below. First, assign initial values for $\{\lambda, \boldsymbol\beta, \bar{F}, \boldsymbol\Delta \}$. Next, iterate the following three steps until the convergence is achieved: (i) for fixed $\lambda$ and $\boldsymbol\beta$, update $\{\bar{F}, \boldsymbol\Delta \}$ via the following estimates from first-order conditions
\begin{align}
    \hat{\bar{F}}= \frac{1}{JT}\sum_{i,j,t} c_i\mathcal{F}_\lambda\left[y_i^j[t] - \mu_\beta(x^j[t],\theta_i)\right],
    \quad
    \hat{\Delta}_i^j=\frac{1}{T}\sum_{t=1}^T|y^j_i[t]-\mu_{\beta}(x^j[t],\theta_i)|_1;
\end{align}
(ii) for fixed $\boldsymbol\beta$ and $\bar{F}$, update the transformation parameter $\lambda$ ignoring the two penalization terms;
and (iii) for fixed $\lambda$, $\bar{F}$ and $\boldsymbol\Delta$, optimize for $\boldsymbol\beta$ via numerical line search methods, e.g., L-BFGS algorithm \citep{liu1989limited}.
The full optimization procedure is provided in Algorithm \ref{alg:BCD}. Since \eqref{eq:paraestimate} is a non-convex optimization problem, the proposed BCD algorithm only converges to a stationary solution \citep{tseng2001convergence}. Because of this, we suggest performing multiple runs of Algorithm \ref{alg:BCD} with random initializations for each run, then taking the converged estimates for the run with smallest objective function. These runs should be performed in parallel if possible, to take advantage of the parallel computing capabilities in many computing systems.

It is important to note the difference between the \textit{training} stage in our calibration-free method and the \textit{calibration} stage in the standard calibration problem \citep{KO2001}.
In calibration methods, the calibration parameter $\eta$, assumed to be a constant, is directly estimated from the training set. This can be viewed as estimating a population average of the historical $\{\eta^j\}_{j=1}^J$, which would \textit{not} be helpful in our cell manufacturing application. Due to the patient-to-patient variability, the calibration parameter $\eta^*$ corresponding to the new patient can be completely different from the historical average value. In contrast, our calibration-free method adopts a patient-invariance statistic $F$, constructed from multiple sensor readings, to alleviate this patient-to-patient variability. In our training setup, we learn the unknown mean relationship $\mu_i(\cdot)$ and the transformation $\mathcal{F}[\cdot]$, which provide the best patient-invariance statistic $F$.  We can then use the invariance statistic $F$ to effectively recover VCCs via \eqref{eq:estimatex}, free from the patient-specific calibration parameter $\eta^*$.

\section{Simulation study} \label{sec:simstudy}

A detailed simulation study is conducted in this section. We first look into a toy application of recovering gravitational acceleration coefficients, to show the applicability of the proposed method.
We then discuss more simulation experiments with different sensing relationships.

\subsection{A gravity application} \label{sec:gapp}
\begin{figure}[t]
\centering
  \includegraphics[width=0.3\linewidth]{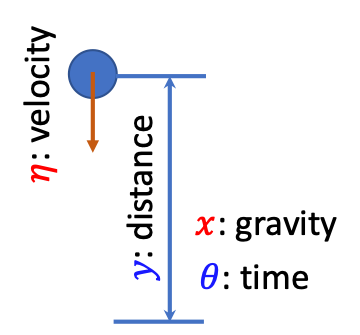}
  \caption{\label{fig:GExample} An illustration and notations of the toy application of recovering the gravitational acceleration coefficient.}
  \end{figure}

Consider the following toy application, where the goal is to recover the gravitational acceleration coefficient $x$, for a different planet. As illustrated in Figure \ref{fig:GExample}, we drop a ball and measure the traveling distance $y$ of the ball after a certain period of time $\theta$. From the physical knowledge, we have the relationship
\begin{align}
    y = f(x,\theta,\eta) = \frac{1}{2}x\theta^2 +\eta\theta.
    \label{eq:illexample}
\end{align}
Here, $y$ is the traveling distance measured by, e.g., taking a photo, $\eta$ is the initial velocity of the ball, and $\theta$ is the time period between dropping the ball and taking the photo. Suppose the ball is dropped by an engineer, meaning that the initial velocity $\eta$ is non-zero and changes among different drops. With the collected data  $\{y,\theta\}$, typically, one cannot recover the gravitational acceleration $x$ even with the known relationship \eqref{eq:illexample}. This is because the initial velocity $\eta$ is also \textit{unknown}. 
The key idea of the proposed calibration-free method is to take \textit{multiple} photos at different time $\theta_i; i=1,\cdots, I$. Therefore, more data $\{y_i,\theta_i\}_{i=1}^I$ is collected with the \textit{same} initial velocity $\eta$. We can then use the proposed invariance statistic and Algorithm \ref{BCD} to ``cancel out" $\eta$ and conduct inference on the gravitational acceleration $x$ of interest.

The setup for recovering the gravity coefficient is as follows. We set the number of photos $I=3$, with parameters  (i.e., the time of taking photos) $\{\theta_i\}_{i=1}^3=\{0.5,1,1.5\}$. 
A historical dataset $\{y_i^j, x^j; \eta^j\}_{j=1}^{J}$ of size $J=100$ is generated with calibration parameters (i.e., initial speed, \textit{unknown}) $\eta^j \sim Unif(1,3)$, gravity coefficients $x^j \sim \mathcal{N}(9.8,1^2)$, and each sensor reading (i.e., traveling distance) $y_i^j$ simulated by \eqref{eq:illexample} with an additional i.i.d. measurement error following $\mathcal{N}(0,0.4^2)$.
To test the recovery accuracy,  we let 
the underlying truth $x^*=9.8$, $\eta^* \sim  Unif(1,3)$ randomly generated, and $y^*$ obtained by \eqref{eq:illexample} with the same measurement error.

The proposed calibration-free method (via Algorithm \ref{BCD}) is applied to find the best transformation $\mathcal{F}_{\hat\lambda} [\cdot]$ and then recover the gravitational acceleration $\hat{x}^*$. The linear combination coefficients are $\{c_i\}_{i=1}^3=\{1,1,-2\}$, and the set of candidate basis functions is $\Phi = \{x,\theta,\theta x, \theta x^2, \theta^2 x\}$. The proposed calibration-free method is repeated, with newly generated test data $\{y^*,x^*=9.8;\eta^*\}$, for 20 times.

We consider the following two baseline methods (also see Table \ref{ta:diffMethods}). First, we implement the supervised learning setting \citep{bishop2006pattern}, i.e., assuming the calibration parameter $\eta = \eta^j = \eta^*$ is the \textit{same} in both training stage and monitoring stage. Such an assumption is \textit{not} true in the considered cell manufacturing application. 
Specifically, we use the historical dataset $\{ y_i^j,x^j\}_{j=1}^{100}$ to estimate an $\bar{\eta}$ and a relationship $g(x,\theta) := f(x,\theta;\bar{\eta})$, which would then be used to recover $\hat{x}^*$. Here, we use the same set of basis functions $\Phi$ for $g(\cdot)$, and a similar optimization scheme with LASSO type penalization \citep{tibshirani1996regression} for estimating the coefficients $\boldsymbol\beta$. This method is referred to as ``SameCal".  The other  baseline method is the standard $l_2$ calibration method suggested by \cite{tuo2015efficient}. In order to adopt this method, we need to assume that the calibration parameters $\{\eta^j\}_{j=1}^{100}$ in the historical data are \textit{known}, which is \textit{not} true in reality. Therefore, we refer this method as ``Oracle". To estimate the sensing relationship $f(\cdot)$, we adopt the set of basis functions $\Phi_o = \{x,\theta,\theta x, \theta x^2, \theta^2 x, \eta x ,\eta \theta, \eta x \theta \}$ and a similar optimization scheme with LASSO type penalization. 
Both baseline methods are implemented to recover $x^*$ of interest via minimizing the squared difference similar to \eqref{eq:estimatex}, using the same 20 simulated test data.

\begin{figure}[]
\centering
      \includegraphics[width=0.85\linewidth]{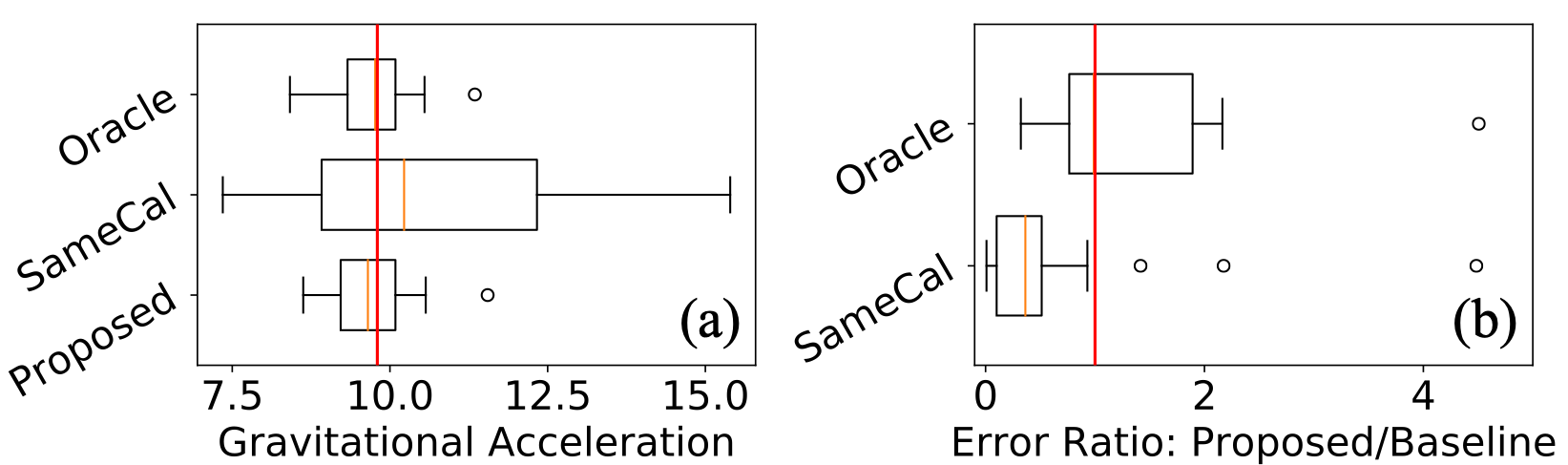}
      \caption{\label{fig:Gerror} Results of the gravity application: (a) shows the recovered gravitational acceleration by the three considered methods. The red line marks the underlying truth $x^*=9.8$. (b) shows the boxplots of absolute error ratios between the proposed method and baseline baseline methods. The red line marks the ratio of 1.0.}
  \end{figure}

Figure \ref{fig:Gerror} (a) shows the boxplots of the estimated $\hat{x}^*$ using the proposed calibration-free method and the two baseline methods. The red line indicates the ground truth value $x^*=9.8$. Among the three methods, the Oracle baseline preforms the best since it queries  \textit{additional} information of the calibration parameter in the training stage, which is again \textit{not} feasible in reality. We notice that the proposed calibration-free method performs almost as good as Oracle. It can accurately recover the true value, with the mean over the 20 estimates $9.7$ and a relatively small variance, whereas for the SameCal baseline, the mean for 20 estimates is $10.8$ and a noticeable bigger variance is observed.

We also conduct a pairwise comparison over the 20 test repetitions. Figure \ref{fig:Gerror} (b) shows the boxplots of absolute error ratios of the proposed method over the two baseline methods, with the red line indicating the ratio of 1.0. We notice that the proposed method is only slightly worse than Oracle; this is impressive since our method do \textit{not} query the underlying calibration parameter in the training stage.
Furthermore, the proposed calibration-free method is noticeably better in recovering the true $x^*$ compared to SameCal. More specifically, our method outperforms SameCal with smaller errors in 17 estimates over 20 test runs in total. This is not surprising since the calibration-free method, utilizing the patient-invariance statistic, can address the patient-specific calibration parameters $\eta^*$.

\subsection{More experiments}

Here, we conduct more experiments on the proposed calibration-free method. Specifically, we consider the following four underlying sensing relationships $f_k(x,\theta,\eta)$:
\begin{enumerate}
\item  $f_1(x,\theta,\eta) = x\theta+\eta\theta^2$;     
\item $f_2(x,\theta,\eta) =  3x+2\theta x+x\theta\eta$;
    \item $f_3(x,\theta,\eta) = \theta x +x^2+\theta\eta^2+\sqrt{\theta}\eta^2+\sqrt{x}\eta/4$;
     \item $f_4(x,\theta,\eta) = \sin(x)+(x+\eta)^\theta+\frac{x}{\theta+\eta}$.
\end{enumerate}
Note that function $f_1(\cdot)$ is quite similar to the sensing relationship \eqref{eq:illexample} in the gravity application in Section \ref{sec:gapp}. For functions  $f_2(\cdot)$ and $f_3(\cdot)$, we notice the existence of interaction terms between $x$ and $\eta$, which means the separable assumption $\delta_i(x,\eta) = \delta_i(\eta)$ in Section \ref{sec:invarstat} does not hold. However, $f_2(\cdot)$ and $f_3(\cdot)$ can still be approximately represented by the adopted set of basis functions $\Phi$. For function $f_4(\cdot)$, it is quite complex, and cannot be represented by $\Phi$. We test all four functions, using the proposed calibration-free method and the same two baseline methods -- SameCal and Oracle -- introduced in Section \ref{sec:gapp}. The detailed test procedure is the same as that in Section \ref{sec:gapp}.

\begin{figure}
\centering
\includegraphics[width=0.8\textwidth]{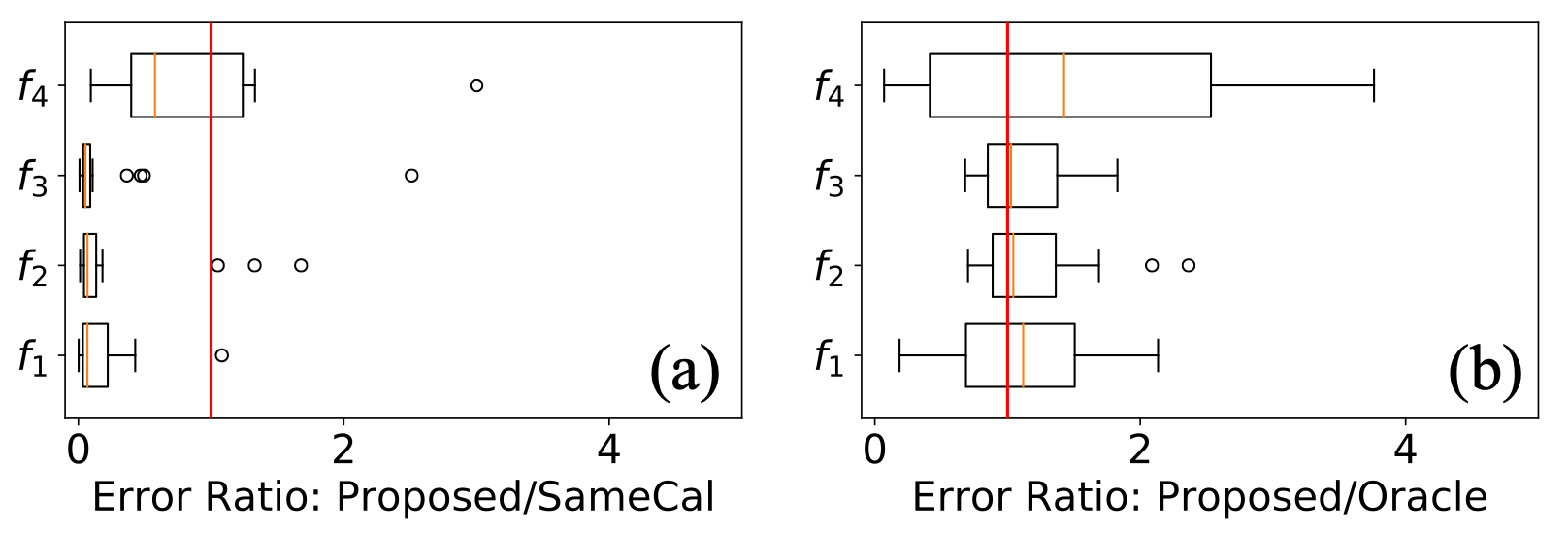}
\caption{\label{fig:simulation} Boxplots of error ratios between the proposed method and the considered baselines, under different sensing relationships. The red line marks the ratio of 1.0, indicating the baseline method achieves the same accuracy as the proposed method.}
\end{figure}

Figure \ref{fig:simulation} shows the boxplots of the absolute error ratios of the proposed method over the baseline SameCal method (a) and the baseline Oracle method (b), under all four underlying sensing functions $f_k(x,\theta,\eta); \; k=1,\cdots, 4$. 
We notice the error ratios in Figure \ref{fig:simulation}  (a) are mostly smaller than $1.0$, indicting that the proposed method can achieve smaller errors compared to SameCal. This is because the assumption of constant calibration parameter in SameCal does \textit{not} hold in cell manufacturing application (and thereby in this simulation study), whereas, our calibration-free method can address the patient-specific calibration parameter via 
a proper combination of multiple sensor readings.
Moreover, compared to Oracle, the proposed method is only slightly worse. This shows the good performance of our calibration-free method: Though we do \textit{not} know the values of the calibration parameter, we \textit{can} recover the underlying parameter of interest similar to the Oracle baseline, where whose values \textit{are} assumed accessible.
Finally, we notice that for the sensing relationship $f_4(\cdot)$, while the proposed method adopts an inappropriate basis decomposition $\Phi$, it still outperforms SameCal.
This is again because the proposed calibration-free method introduces the invariance statistic to alleviate the effect of patient-to-patient variability, and therefore, shows improved performances in recovering the quantity of interest.

\section{Cell manufacturing case study} \label{sec:casestudy}

In this section, we apply the proposed calibration-free method to the motivating case study of cell manufacturing. As discussed in Sections \ref{sec:intro} and \ref{sec:probdefin}, we are interested in recovering and monitoring the Viable Cell Concentration (VCC) $x[t]$ at different experimental time $t$. This is because the goal of cell manufacturing is to culture a small batch of cells to a significant amount, for an effective re-administering in the CAR T cell therapy (see Figure \ref{fig:CART}). 

\subsection{Experimental setup}
In our experiment, human leukemic T cells (Jurkat E6-1; American Type Culture Collection, ATCC\textsuperscript{\tiny\textregistered}) are cultured in ATCC-formulated culture medium (RPMI-1640; GE Healthcare) with 10\% fetal bovine serum, 2 mM L-glutamine, 10 mM 4-(2-hydroxyethyl)-1-piperazineethanesulfonic acid (HEPES), 1 mM sodium pyruvate, 4500 mg/L glucose, and 1500 mg/L sodium bicarbonate in a 75 cm\textsuperscript{2} petri dish (Nunc\texttrademark~
EasYFlask\texttrademark; ThermoFisher Scientific\texttrademark). The cells are cultured in a humidified incubator controlled at 37\textdegree{}C and 5\% CO\textsuperscript{2}, and the culture media is pre-heated to avoid the temperature effect on the impedance measurement \citep{dlugolkecki2010resistances}.

\begin{figure}
\centering
\includegraphics[width=0.85\textwidth]{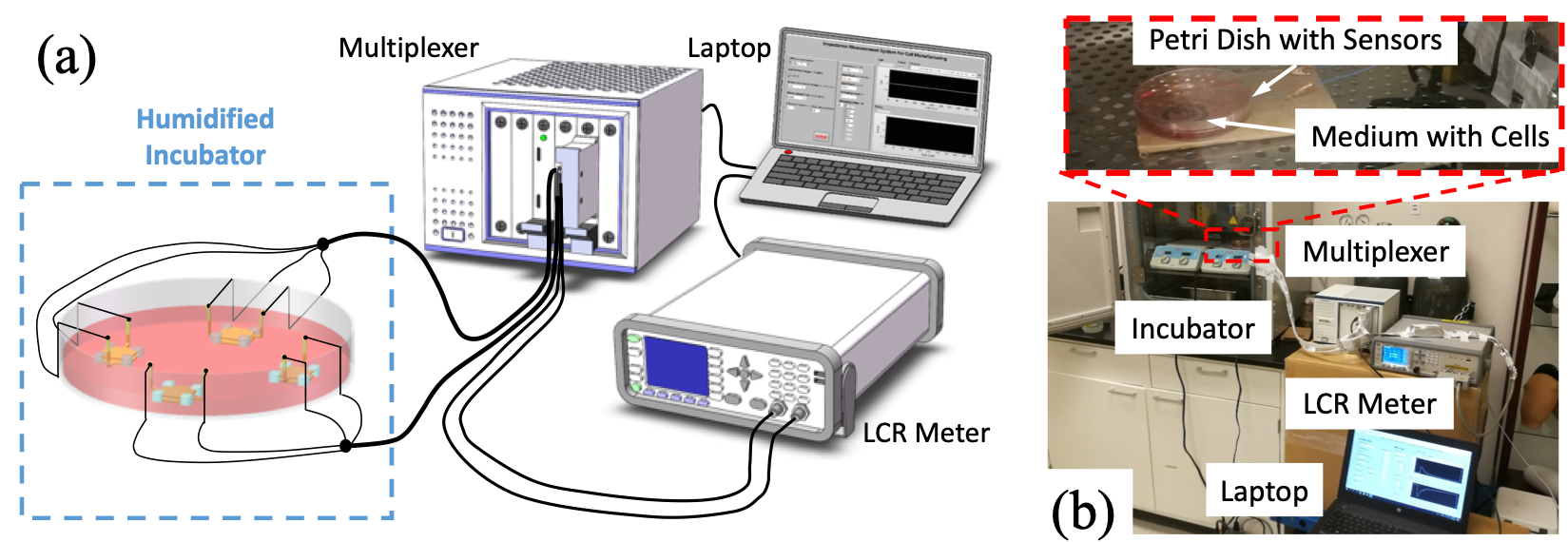}
\caption{\label{fig:ExpSetup} The cell manufacturing application: (a) an illustration and (b) the actual experimental setup with an emphasis on the impedance measurement part.}
\end{figure}

The impedance measurements are obtained by our electric cell-substrate impedance sensing. Figure \ref{fig:ExpSetup} illustrates the experimental setting for the impedance measurement. Here, we use $I=4$ sandwich shape 3D impedance sensors, consisting a pair of parallel-plate electrodes and PDMS (Sylgard 184, Tow Corning) to maintain a gap between two electrodes (see Figure \ref{fig:biosensor} (b)). In our experimental setup, the geometry parameter $\theta$ of the sensors is the edge length of the electrode pads, and $\{\theta_i\}_{i=1}^4 = \{8 mm, 10 mm, 12 mm, 14 mm\}$. Impedance measurements are conducted by an LCR meter (E4980AL; Keysight Technologies) with a sinusoidal signal of 22 mVrms under multiple frequencies ranging from 500 Hz to 100 kHz. We let impedance measurement $y$ be the relaxation strength computed from the raw impedance readings over frequencies, i.e., the difference between permittivity values of high-frequency end and low-frequency end of the dielectric relaxation process, for its high dependency to the VCC of interest \citep{schwan1957electrical}.
The measurement is taken every $15$ minutes for around 30-35 hours. This is because typically after 35 hours, we have to change the culture media and expand the cells to a bigger cell culture flask, which would inevitably interrupt the online monitoring. This results in an online monitoring dataset $\mathcal{D}_{\rm monitor}={\{y_i^j[t],\theta\}_{t=1}^T}_{j=1}^J$ with the underlying VCCs to be recovered.

The ground truth VCCs are obtained by an automated cell counter (TC20\texttrademark; Bio-Rad Laboratories, Inc.), and the concentration is maintained between 1 × 10\textsuperscript{5} and 1 × 10\textsuperscript{6} cells/mL. Multiple repetitions are performed, with the averaged value reported as the underlying VCC $x$. Note that the measurement procedure is not only labor-intensive but may also introduce contamination to the culture media (see Section \ref{sec:intro}). We will only measure VCC around six times per cell culture experiment, which leads to a much smaller ($\tau \ll T$) yet full training dataset $\mathcal{D}_{\rm train}={\{y_i^j[t],x^j[t],\theta\}_{t=1}^{\tau}}_{j=1}^J$.

We conduct the cell culturing for $J=5$ experiments, each with different starting materials, i.e., different calibration parameters $\eta^j$. We use the same $I=4$ sensors $\{\theta_i\}_{i=1}^4 = \{8 mm, 10 mm, 12 mm, 14 mm\}$ for the $J=5$ experiments.
In experiment $j$, we measure and compute the relaxation strength of impedance $y^j_i[t]$ for each sensor $i$, at different time $t$ (every 15 minutes, $T\approx 130$ in total).
Meanwhile, we measure the ground truth VCC $x^j[t]$ with a much lower resolution ($\tau \approx 6$ in total).

\begin{table}[t]
\centering
\begin{tabular}{l|ccccc|c}
 \toprule
         & \textbf{Exp 1} & \textbf{Exp 2} & \textbf{Exp 3} & \textbf{Exp 4} & \textbf{Exp 5} & \textbf{Mean}\\
       \hline
\textit{Proposed} &   \textbf{0.092}    &  0.270      &   \textbf{0.080}  &      \textbf{0.379}   &    \textbf{0.590}  & \textbf{0.282}   \\
\textit{SameCal} &    0.500   &  \textbf{0.174}     &   0.328      &   0.742      & 0.760  &0.501  \\ 
 \toprule
\end{tabular}
\caption{Cross-validation errors of the recovered VCCs for the cell manufacturing case study, using the proposed calibration-free method and the baseline SameCal  method.}
\label{table:cell}
\end{table}

\subsection{Cross validation of viable cell concentration}

For the collected training dataset $\mathcal{D}_{\rm train}$, we first preform a cross-validation test \citep{friedman2001elements} on the recovered VCCs. 
More specifically, 
we apply the proposed calibration-free method (via Algorithm \ref{BCD}) to learn the sensing relationship using four out of five experiments, and then recover VCC $\hat{x}^j[t]$ for the remaining experiment via \eqref{eq:estimatex}.
We let the linear combination coefficients $\{c_i\}_{i=1}^4 = \{1,-1,1,-1\}$ and select the same set of basis functions $\Phi$ as in Section \ref{sec:gapp}. Furthermore, a log-transformation on VCCs is performed prior to the analysis. 
We consider SameCal (see Section \ref{sec:simstudy}) as the baseline method. Such a method introduces an additional assumption that the calibration parameter is a constant. Note that the Oracle baseline cannot be adopted here since the actual values of the calibration parameter are always \textit{unknown}, which is the key motivation of the proposed calibration-free method (also see Section \ref{sec:p2pvar} and Table \ref{ta:diffMethods}). 

Table \ref{table:cell} shows the absolute errors of the recovered log VCCs when the ground truth VCCs are measured in each experiment. We observe that the proposed method outperforms the baseline SameCal method four experiments out of five. Furthermore, the mean error of the five experiments by the proposed method is $0.282$, which is almost two times smaller than that of $0.501$ by SameCal. This is due to the fact that the calibration parameter, which models the patient-to-patient variability, is \textit{not} a constant in cell manufacturing \citep{hinrichs2013reassessing}; the proposed calibration-free method properly addresses this variability via the construction of the patient-invariance statistic.

\subsection{Online recovery of viable cell concentration}

\begin{figure}
\centering
\includegraphics[width=0.9\textwidth]{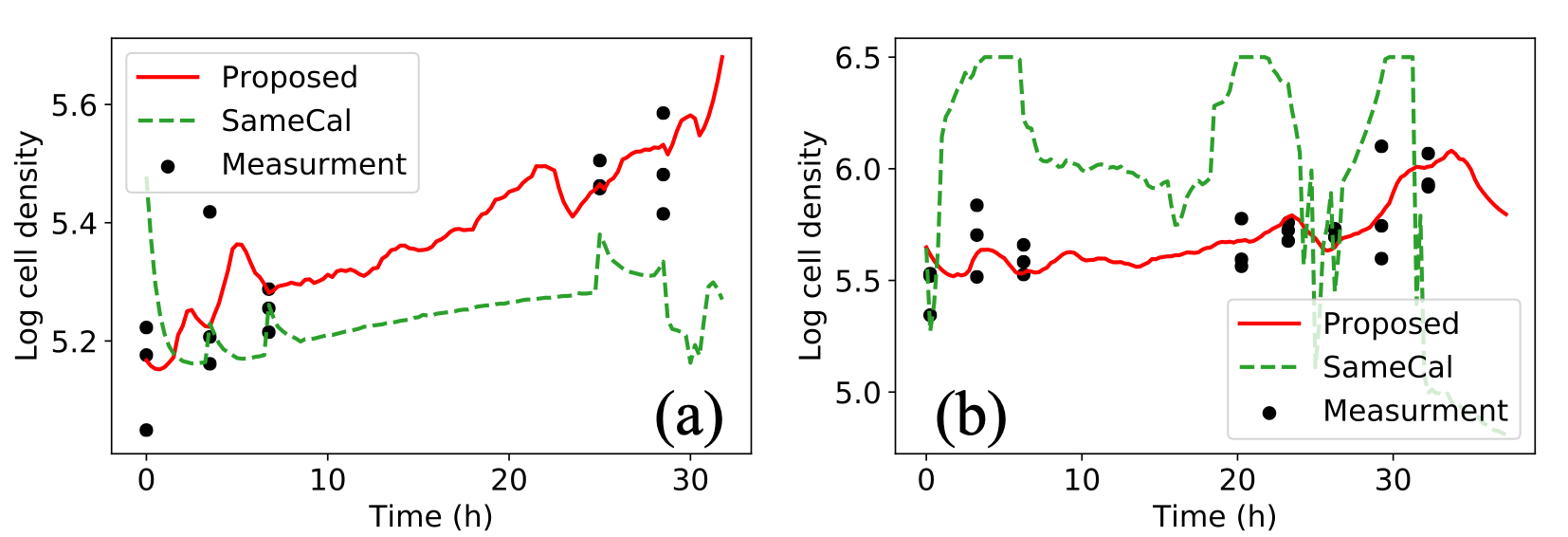}
\caption{\label{fig:CellDensity} The recovered VCC over time of two cell manufacturing experiments under the two considered methods. The ground truth VCC measurements are shown in dots.}
\end{figure}

We then perform VCC recovery on the online monitoring set $\mathcal{D}_{\rm monitor}$. Here, the sensing relationship is estimated using all five experiments in the training set $\mathcal{D}_{\rm train}$. Figure \ref{fig:CellDensity} shows the two recovered log VCC curves over the whole culture time $\log \hat{x}^j[t]$, via the proposed calibration-free method (in red line) and the baseline SameCal method (in green dash line). The ground truth (log) VCC measurements in $\mathcal{D}_{\rm train}$ are also plotted in black dots. 
We see that the proposed method recovers a meaningful estimation of VCC. The recovered $\log \hat{x}^j[t]$ increases approximately linear over the culture time $t$, indicating $\hat{x}^j[t]$ growths  exponentially in time; this matches the preliminary understanding in the cell culture literature \citep{haycock20113d}. Furthermore, the recovered $\hat{x}^j[t]$ approximately passes through the ground truth measurements. However, due to the huge patient-to-patient variability, the baseline SameCal method struggles in either passing through the ground truth experiments or providing reasonable estimates of VCC curves. Our calibration-free method, adopting the patient-invariance statistic, appears to alleviate such variability well.

The proposed calibration-free method can also provide important biological insights for cell growth in cell manufacturing. From Figure \ref{fig:CellDensity} (b), we notice a decrease in the VCC curve at around hour 32. This may be due to the lack of nutrition in the media since the culture media typically needs to be changed after 30 hours. 
Furthermore, we observe from Figure \ref{fig:CellDensity} that the VCC curves decrease slightly in the first two hours in cell manufacturing. This may be because of the lack of viability of the cells at the beginning of the culture process -- 
though we have already thaw cells and stood them still for several minutes, it seems that a certain portion of cells still do not gain full viability and die soon. 
As a result, we suggest standing the cell still longer for future experiments.
Last but not least, we notice a small VCC decrease when conducting the ground truth VCC measurements. One reason for this is that the measurement itself is not in-line and needs to contact the culture media; it may introduce contamination, and therefore, kill a small portion of cells \citep{haycock20113d}. In contrast, the proposed calibration-free biosensing method, together with impedance-based biosensors, provides an in-line, non-destructive, and non-contact way for VCC monitoring in cell manufacturing.

\section{Conclusion} \label{sec:conclusion}
In this work, we propose a new calibration-free method for monitoring viable cell concentration in cell manufacturing, which is a critical component in the promising CAR T cell therapy.
The key challenge here is the patient-to-patient variability in the initial culturing material, leading to poor performances in recovering viable cell concentrations via existing methods.
We propose to use multiple impedance-based biosensors with different geometries and an associated calibration-free statistical framework  for online recovery of viable cell concentrations.
Specifically, we model the patient-to-patient variability via a patient-specific calibration parameter. We then construct a patient-invariance statistic, which uses a transformation and a linear combination of sensor readings to alleviate the effect of the calibration parameter. In the training stage, we learn the best transformation and the sensing relationship via a carefully formulated optimization problem. In the online monitoring stage, viable cell concentrations can be recovered via the invariance statistic, free from the patient-specific calibration parameter. We then apply the proposed calibration-free method in different simulation experiments and a real-world case study of cell manufacturing, where the proposed method demonstrates substantial improvements against the existing methods. Therefore, we believe the proposed calibration-free method can play an essential role in cell manufacturing and reduce the cost of the promising CAR T cell therapy.

Looking ahead, there are several interesting directions for future exploration. To begin with, a more thorough analysis of impedance-based sensors can be conducted, with a detailed comparison of sensitivity using different experimental settings such as sensor geometries and electrode materials. 
Moreover, we adopt in this work a parametric sensing relationship and a heuristic approach for parameter estimation. This is mainly due to the already improved performance compared to the baseline methods. A more flexible, and non-parametric Gaussian process regression method \citep{santner2013design,lin2019transformation} with a rigorous likelihood-based parameter estimation scheme may lead to further improvements in recovering viable cell concentrations, as well as other critical quality attributes.
Finally, micro cameras can also be used in cell manufacturing. Therefore, we are also interested in monitoring cell manufacturing based on cell morphology. In this case, physics-informed deep learning frameworks in the literature \citep{raissi2017physics,chen2020active} appear to be suitable for recovering critical quality attributes in cell manufacturing.



\end{document}